\documentclass[aps,pra,reprint]{revtex4-1}

\usepackage{graphicx}
\usepackage{bm}
\usepackage{bbold}
\usepackage{amssymb}
\usepackage{amsmath}
\usepackage{natbib}

\usepackage{color}

\renewcommand{\vec}[1]{{\mathbf #1}}

\bibpunct{[}{]}{,}{n}{}{}

\begin{document}

\title{The influence of evanescent waves on the nature of optical cooperative effects in atomic ensembles in a waveguide}

\author{A. S. Kuraptsev}
\email[]{aleksej-kurapcev@yandex.ru}
\affiliation{Peter the Great St. Petersburg Polytechnic University, 195251, St. Petersburg, Russia}
\affiliation{\small Institute for Analytical Instrumentation of RAS, 198095, St. Petersburg, Russia}

\author{I. M. Sokolov}
\email[]{igor.m.sokolov@gmail.com}
\affiliation{Ioffe Institute, 194021, St. Petersburg, Russia}

\date{\today}

\begin{abstract}
Based on a consistent quantum microscopic approach, we investigate the peculiarities of collective polyatomic effects in atomic ensembles placed in a waveguide, caused by the presence of
evanescent modes of electromagnetic field. We analyze the influence of these modes on the process of cooperative spontaneous decay,
as well as on the nature of radiation transfer in the ensembles under consideration.
We show that under certain conditions, their influence can be dominant compared to the role of
radiation modes, and the mechanism for this influence is the modification of dipole-dipole
interatomic interaction.
\end{abstract}

\maketitle

\section{Introduction}
Ensembles of point quantum emitters (atoms, ions, quantum dots) embedded in a transparent dielectric are considered as promising objects for a wide range of applications in quantum electronics, nanophotonics and quantum information science \cite{1,2,3,4,5,6,7}. Such ensembles exhibit high spectral selectivity as well as sensitivity to external influences. Depending on the specific application, different methods are used to improve the efficiency of their use. In a wide number of problems, optical cavities and waveguides are used to enhance the interaction of atoms with electromagnetic radiation.

Since the seminal work of Purcell \cite{8}, it is well known that radiative characteristics of an atom in a cavity/waveguide differ from its free space values. The reason of these changes is the modification of the spatial structure of the modes of the electromagnetic field, including vacuum reservoir. In its turn, this modification leads to the alteration of the coupling constant between an atom and electromagnetic field. Up to now, the characteristics of individual quantum emitters, such as the lifetimes of excited states, level shifts and widths, radiation patterns, etc. has been studied in detail in cavity QED (see \cite{9,10,11,12,13,14,15,16, 17,18,19} and references therein).

Beside single-atom characteristics, a cavity also changes the electromagnetic interaction between different particles. In particular, it leads to the modification of interatomic dipole-dipole interaction \cite{20, 21} and associated many-body cooperative effects \cite{13,22,23,24,25,26,27,28}, which have been studied in less detail.

The role of cooperativity is particularly important for atomic ensembles located in a waveguide. The reason of this is that even far-separated atoms are strongly coupled to each other due to long-range radiation waves, which propagate directionally along the axis of a waveguide. In the optical domain, a waveguide is usually represented as a photonic crystal fiber filled with a gas of active atoms or an optical fiber doped with active impurities \cite{Bufetov1}. In the microwave domain, a waveguide is usually represented as a hollow metal tube. Herewith the probe radiation is tuned on the microwave transition between Rydberg states of active atoms \cite{Abmann}. Alternatively, in the frame of microwave experiments, macroscopic scatterers can be used instead of atoms \cite{Chabanov}. Random ensembles of point scatterers located in a waveguide attract special attention due to the possibility of achieving the effect of Anderson localization of light \cite{Chabanov,30,31,32,33,34}.

When studying the optical properties of atomic ensembles in a waveguide, we have recently shown that changing the sizes of cross section of the waveguide significantly affects the nature of radiation transport \cite{34}. With the increase of the transverse sizes, the transition from the regime of Anderson localization of light to traditional diffusive radiation transfer is observed. It is connected with the number of radiation modes, which are able to propagate over long distances inside the waveguide as traveling waves.

In addition to oscillating radiation modes, exponentially decaying evanescent modes of the electromagnetic field are also present in a waveguide. The main goal of this work is to analyze the influence of these modes on the nature of collective polyatomic optical effects. We will show that in some cases these modes significantly affect the interatomic dipole-dipole interaction even for far-separated atoms. In its turn, this alters the lifetimes of the collective states of the atomic ensemble, the dynamics of cooperative spontaneous decay, and, subsequently, the radiation transport.

The study of the role of evanescent modes is particularly interesting because it helps to better understand the unexpected dependence of the transmission coefficient on the transverse size of the waveguide, which was shown in Ref. \cite{34}, Fig. 4. It is especially important in the vicinities of the critical points of the transverse size, which correspond to the changes of the number of radiation modes. Near these critical points, extremely sharp dependence was observed, and its explanation was not comprehensively understood so far. Furthermore, evanescent modes are of great interest because they play a key role in a number of experiments with cold atoms trapped near a nanofiber, see for example Refs. \cite{nanofiber1,nanofiber2}.

\section{Basic assumption and highlights of the approach}
\begin{figure}\center
	\includegraphics[width=7cm]{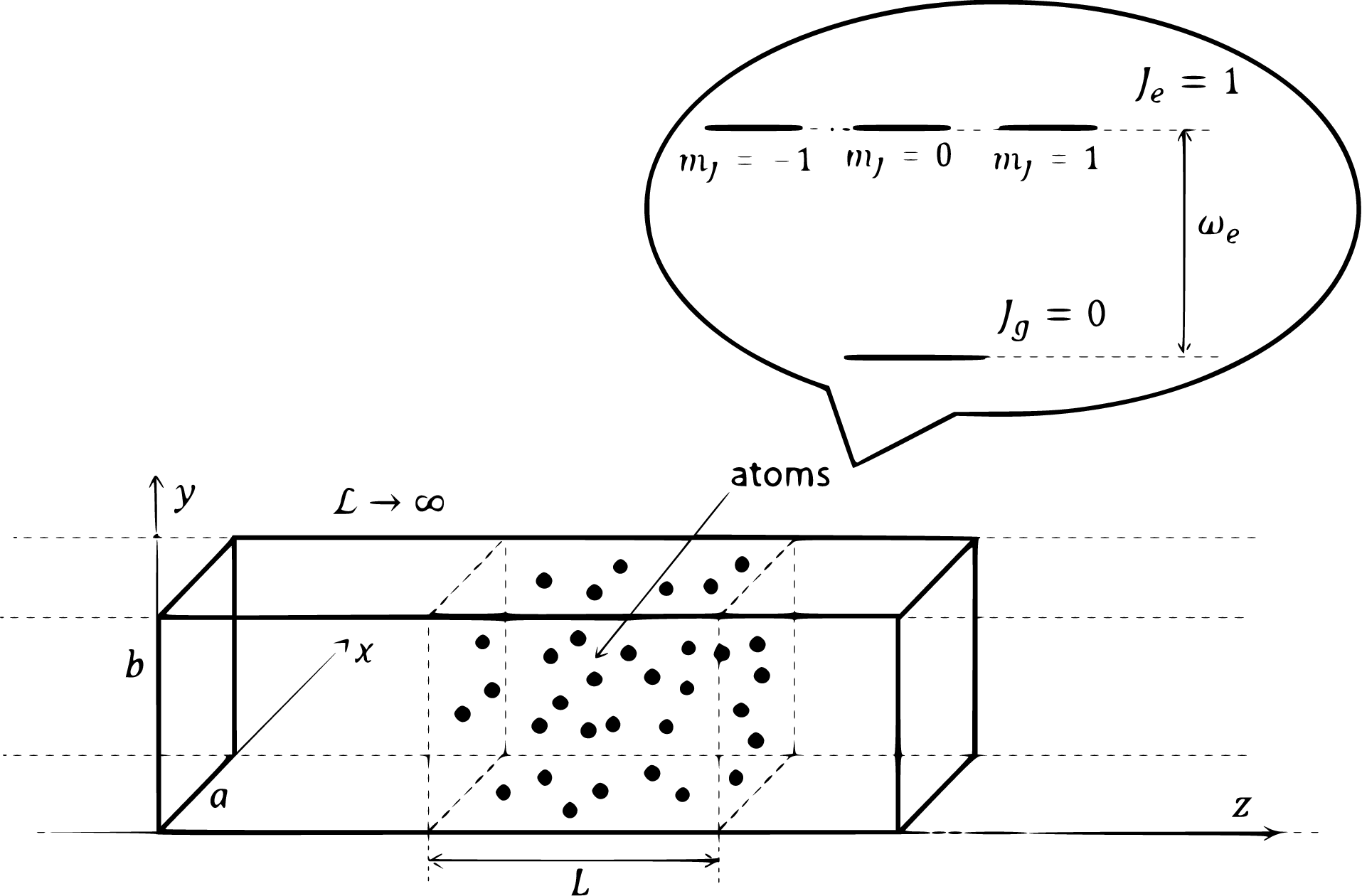}
\caption{\label{fig:one}
 Sketch of the waveguide and the atomic ensemble inside it. The inset
illustrates the model scheme of atomic energy levels.}\label{f1}
\end{figure}

We consider an ensemble of $N$ two-level atoms at random positions $\{ \vec{r}_i \}$ inside a waveguide, see Fig. \ref{f1}. This model is well suited to describe impurity atoms embedded in a solid transparent dielectric at low temperatures \cite{Naumov2}. Each atom has the ground state $|g\rangle$ with the energy $E_g$ and angular momentum $J_g = 0$; as well as triply degenerate excited state $|e\rangle$ with the angular momentum $J_e = 1$ and the energy $E_e = E_g + \hbar\omega_e$. Three Zeeman sublevels of the excited state are characterized by the projection of the angular momentum on the quantization axis $z$ -- $m_J = 0,\pm1$. The free space natural linewidth of each Zeeman sublevel of the excited state is $\gamma_0$.

We will assume that the atoms are motionless. We will neglect the effects associated with thermal vibrations of the atoms \cite{Skipetrov2025}, considering the case of low temperatures and assuming that the spectral lines of impurity atoms are zero-phonon \cite{Rebane}.

When describing the waveguide, we will not assume its quasi-one-dimensionality. The waveguide's shape and the spatial and polarization structure of its modes will be accurately accounted for. For simplicity, we will consider a waveguide of rectangular cross-section with sides $a$ and $b$. Its walls are assumed to be perfectly conducting. We will neglect the absorption of waves during their propagation. We also assume that the excitation of the ensemble is so weak that all nonlinear optical effects can be neglected.

The main difficulty in describing collective effects is the need to account for so-called recurrent scattering, where the same photon can be scattered multiple times by the same atom as a result of rescattering within the medium. Under these conditions, the coupled dipole (CD) method has proven its effectiveness.

This method was proposed by Foldy \cite{F45} and afterwards discussed at length by Lax \cite{L51}. Later, similar approach was applied to analyze different aspects of collective effects such as collective spontaneous decay, single-photon superradiance and subradiance, strong (Anderson) localization of light and so on \cite{Javanainen:1999,RMO00,34a,Svidzinsky:2010,KSH11,38a,SS14,BGAK14,KS_PRA2015,Guerin:2016a,31a,Guerin:2017a,Kuraptsev:2017,Skipetrov_2019b}.

For the analysis of collective effects in a waveguide, the general concept of the CD method was adapted in our previous works \cite{13,34,35} and we will not reproduce it in detail here. In the following paragraphs, we present just a brief overview of the theoretical developments.

General formalism allows us to consider both time-dependent problems and steady-state ones. Let's start from dynamic problems.
With the approximations described above, we can write the non-steady-state Schrodinger equation for the wave function $\psi$ of the joint system, which consists of the atoms and the electromagnetic field in a waveguide, including a vacuum reservoir. This system is described by the following Hamiltonian:
\begin{eqnarray}\label{equation1}
  \widehat{H}=\sum_{i=1}^{N}\sum_{m_J=-1}^{1}E_{e_{i,m_J}}|e_{i,m_J}\rangle\langle e_{i,m_J}| \nonumber \\
+\sum_{\textbf{k},\alpha}\hbar\omega_{k}\left(\widehat{a}_{\textbf{k},\alpha}^{\dagger}\widehat{a}_{\textbf{k},\alpha}+\frac{1}{2}\right)-
\sum_{i=1}^{N}\widehat{\textbf{d}}_{i}
  \cdot\widehat{\textbf{E}}\left(\textbf{r}_{i}\right) ,\label{equation1}
\end{eqnarray}
where the first two terms correspond to noninteracting atoms and the electromagnetic field in an empty waveguide, respectively; the third term describes the interaction between the atoms and the electromagnetic field in the dipole approximation. In Eq. (\ref{equation1}), $\widehat{a}_{\textbf{k},\alpha}^{\dagger}$ and $\widehat{a}_{\textbf{k},\alpha}$ are the
operators of creation and annihilation of a photon in the corresponding mode, $\omega_{k}$ is the photon frequency, $\widehat{\textbf{d}}_{i}$ is the dipole moment operator of the atom $i$, $\textbf{r}_{i}$ is the position of the atom, $\widehat{\textbf{E}}\left(\textbf{r}\right)$ is the electric displacement vector in a waveguide.

The specific form of the operator $\widehat{\textbf{E}}\left(\textbf{r}\right)$ corresponding to the chosen waveguide geometry is found in a standard way by solving Maxwell's equations with given boundary conditions \cite{36} and subsequent canonical quantization of the electromagnetic field \cite{37}.

We look for the wave function $ \psi $ of the considered closed quantum system system in the form of a decomposition over the eigenfunctions $ {\psi_l} $ of the Hamiltonian of noninteracting atoms and light, $ \psi = \sum\limits_l b_l \psi_l $. Then, considering the case of weak excitation and restricting ourselves by the states of the atomic-field system containing no more than one photon, for the amplitudes $ b_e $ of one-fold excited atomic states $ | \psi_e \rangle= | g \cdots e \cdots g \rangle\equiv | e_i\rangle $ we have the following set of equations:
\begin{equation}
\frac{\partial b_e}{\partial t} = \left( i\omega_e-\frac{\gamma_0}{2} \right)b_e + \frac{i\gamma_0}{2} \sum_{e' \neq e} V_{ee'}b_{e'}.
\label{e1}
\end{equation}
Here, the index $ e $ shows the number of the atom which is excited in the state $| \psi_e \rangle = | g \cdots e \cdots g \rangle $, as well as specific populated Zeeman sublevel.

The first term on the right side of Eq. (\ref{e1}) describes free evolution of independent atomic dipoles. The last term in Eq. (\ref{e1}) corresponds to the interatomic dipole-dipole interaction, and it is responsible for all collective effects. In this term, $V_{ee'}$ is the so-called matrix of re-emission, which describes the photon exchange between the atoms. Detailed analytical expression for this matrix in the case of atomic ensemble in a waveguide with corresponding derivation can be found in the Appendix in Ref. \cite{13}. Since final equations for the elements of the matrix $V_{ee'}$ are bulky, we will not reproduce them here.

System of equations (\ref{e1}) must be supplemented with an initial condition. Assuming weak excitation of an atomic ensemble, we will consider the case when only one atom is initially excited and the electromagnetic field is in a vacuum state. System (\ref{e1}) contains 3N equations, and we solve it numerically. For simplicity, we will assume that transition frequency is the same for all atoms forming the ensemble, $\omega_e=\omega_0$.

To analyze the steady-state radiation transfer through the ensemble under consideration, we can use different approaches. For instance, we are able to consider the steady-state solution of the set of equations (\ref{e1}), completing it with the term, which describes the interaction of atoms with steady-state probe field. However, we find it more convenient to use another approach, where the probe radiation is simulated by the spontaneous emission of the so-called source-atom \cite{34}. In the framework of this approach, we apply the formalism of the dynamical problem and assume that initially excited source atom is far separated from the atomic ensemble. It has the same level structure as atoms of the ensemble but different resonant transition frequency, $\omega_s$, and natural linewidth $\gamma_s$. In order to simulate monochromatic probe radiation, we pass to the limit: $\gamma_s \rightarrow 0$. Next, we consider $t \rightarrow \infty$ assuming $\gamma_s t \ll 1$. Thus, after these limiting passages, we obtain stationary quantum amplitudes $b_e$.

The intensity of the radiation emitted by the ensemble can be calculated in a direct way, as it was done in Refs. \cite{Kuraptsev:2017,KSH11}, or it can be alternatively simulated by the consideration of the so-called "atom-detector" \cite{34}. The latter represents an imaginary elusive "atom", which accepts all radiation emitted by the environment medium but does not re-emit photons. The sensitivity of atom-detector does not depend of the polarization of radiation. So it works as a point detector. The intensity of light at the point of atom-detector is proportional to its excited state population. Thus, to determine the transmission coefficient, $T$, it is enough to calculate the population of the excited state of the atom-detector for two cases: when it is located far ahead from ensemble and far behind it.

\section{Results and discussion}

We investigate the influence of exponentially decaying evanescent modes on the nature of collective effects by analyzing how the optical properties of an atomic ensemble change with the increase of the transverse dimensions of the waveguide, retaining the number of radiation modes to be constant. It is well known that radiation modes of the electromagnetic field at the atomic transition frequency must satisfy the condition that follows from the dispersion relation. For a rectangular waveguide, it reads: $(\pi m/a)^2+(\pi n/b)^2\leq(\omega_0/c)^2$. The indices $m$ and $n$ are positive integers for TM modes, and for TE modes $m,n = 0,1,2,\ldots $, herewith both indices cannot be zero together. If the above condition is not matched for given values of the indices $m$ and $n$, it means that corresponding mode is exponentially decaying.

Next in this paper, we will consider two most interesting cases, when the effect of evanescent waves is mostly pronounced. First, we consider a zero-mode waveguide, where the photon exchange between different atoms is possible only due to evanescent field. Zero-mode waveguide is characterized by the conditions $k_0 a<\pi$ together with $k_0 b<\pi$, where $k_0 = \omega_0/c$. For definiteness, we will keep one of the sizes of the cross section to be constant throughout the paper: $k_0 a=3$. Other size, $b$, we will increase from the value $k_0 b=3$ approaching to the critical value $k_0 b=\pi$.

Second, we consider a single-mode waveguide. It corresponds to $k_0 a<\pi$ and $\pi<k_0 b<2\pi$. The case of single-mode waveguide is the most interesting from practical point of view. Atomic ensemble can be excited by external probe wave, and cooperative effects are essential for arbitrarily low atomic density due to strong dipole-dipole coupling. The latter circumstance significantly distinguishes single-mode waveguide from multi-mode one. In this part, we will fix the size $k_0 a=3$ and vary the parameter $b$, bringing it closer to the critical value $k_0 b=2\pi$.

\subsection{Excitation dynamics}
As a proof of concept, let us consider a model system, which consists of two atoms located in a zero-mode waveguide. First atom is initially excited at Zeeman sublevel $m_{J}=-1$, whereas second atom is initially in the ground state. The atoms are far separated, so the interatomic distance is much larger than the wavelength of resonant radiation. Both atoms are located at the center of the cross section of the waveguide. Interatomic distance is $k_{0}(z_2 - z_1)=100$.

\begin{figure}\center
	\includegraphics[width=7cm]{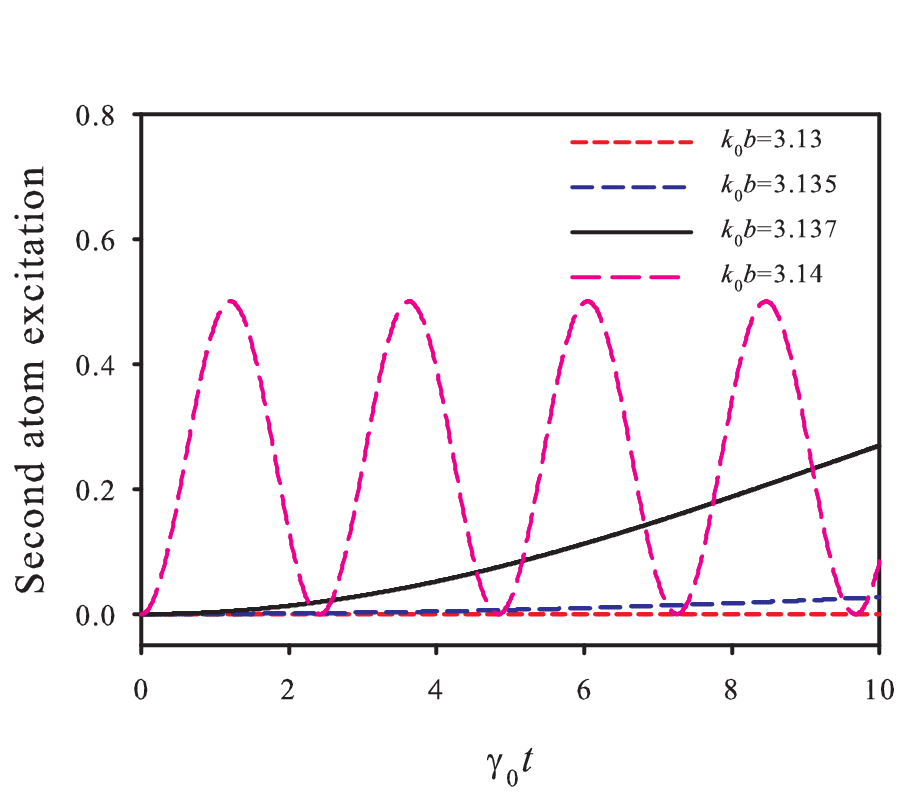}
\caption{\label{fig:two}
 Dynamics of the excited state probability for the second (initially unexcited) atom. $k_0 a=3$. The coordinates of the first (excited) atom are $x_1=a/2$, $y_1=b/2$, $z_1=0$. The coordinates of the second (unexcited) atom are $x_2=a/2$, $y_2=b/2$, $k_{0}z_2=100$. At $t=0$, only one Zeeman sublevel, $m_J = -1$, of the excited state of the first atom is populated.}\label{f2}
\end{figure}

Figure \ref{fig:two} shows the dynamics of the excited state probability of the second atom in the case $k_0 a=3$ for four different values of the parameter $k_0 b$: 3.13, 3.135, 3.137, 3.14. One can clearly see that the increase of the parameter $k_0 b$ approaching its critical value $k_0 b=\pi$ dramatically changes the picture. At the considered time interval, $0\leq \gamma_0 t\leq 10$, the excited state probability of the second atom is:

-- negligibly small when $k_0 b = 3.13$;

-- beginning to manifest itself when $k_0 b = 3.135$;

-- significant when $k_0 b = 3.137$;

-- of fundamental importance when $k_0 b = 3.14$.

Further approaching of the parameter $k_0 b$ to its critical value, $k_0 b=\pi$, leads to the decrease of the period of oscillations observed for $k_0 b = 3.14$. The explanation of the revealed effect is as follows. The mode of the electromagnetic field $\text{TE}_{01}$ formally remains evanescent. But its attenuation constant crucially depends on the parameter $k_0 b$. When it approaches to its critical value, $k_0 b=\pi$, the attenuation constant of $\text{TE}_{01}$ mode limits to infinity. It makes it possible to couple atoms via this mode, even when the interatomic separation is very large.

\begin{figure}\center
	\includegraphics[width=7cm]{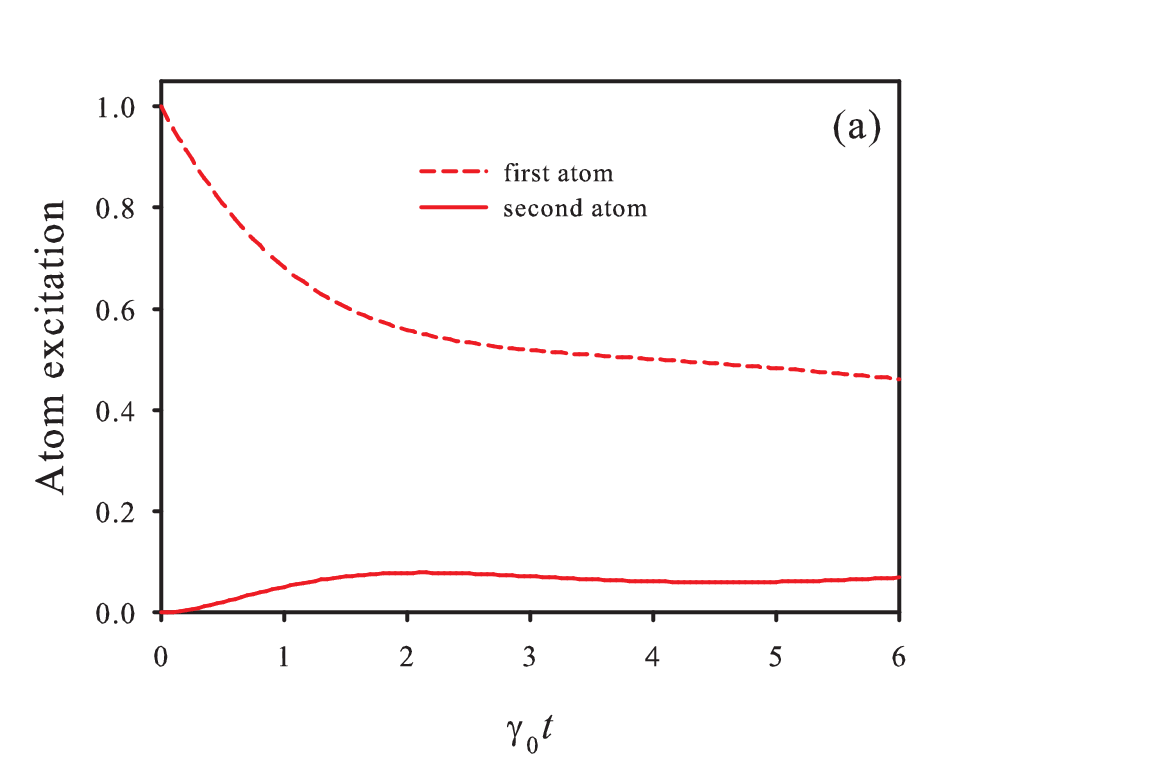}
	\includegraphics[width=7cm]{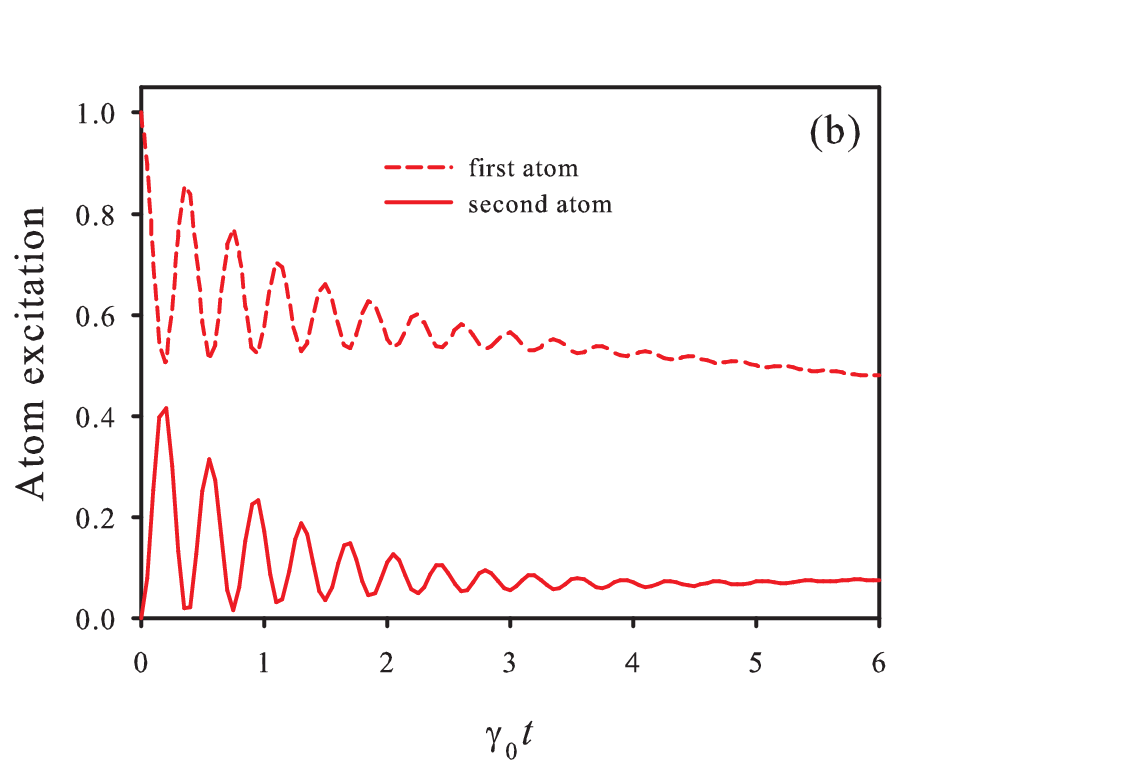}\\
	\caption{\label{fig:three}
			Excitation dynamics for the atoms located in a single-mode waveguide. $k_0 a=3$. $k_0 b=6$ -- (a), $k_0 b=6.28$ -- (b). The coordinates of the first (excited) atom are $k_0 x_1=1$, $k_0 y_1=k_0 b/2-1$, $z_1=0$. The coordinates of the second (unexcited) atom are $k_0 x_2=2$, $k_0 y_2=k_0 b/2+1$, $k_{0}z_2=10$. At $t=0$, only one Zeeman sublevel, $m_J = -1$, of the excited state of the first atom is populated.}\label{f3}
\end{figure}

Next, let us analyze, how the described effect manifests itself in the case of a single-mode waveguide. Fig. \ref{fig:three} demonstrates the excitation dynamics for two cases: when the parameter $k_0 b$ is far from its critical value -- Fig. \ref{fig:three}(a) as well as when $k_0 b$ is close to its critical value -- Fig. \ref{fig:three}(b). In the first case, the photon exchange between the atoms is caused by the radiation mode $\text{TE}_{01}$. In the second case, the attenuation constant of evanescent mode $\text{TE}_{02}$ becomes larger than the interatomic distance, so this mode is largely involved in photon exchange. This results in the oscillations observed in Fig. \ref{fig:three}(b).

\begin{figure}\center
	\includegraphics[width=7cm]{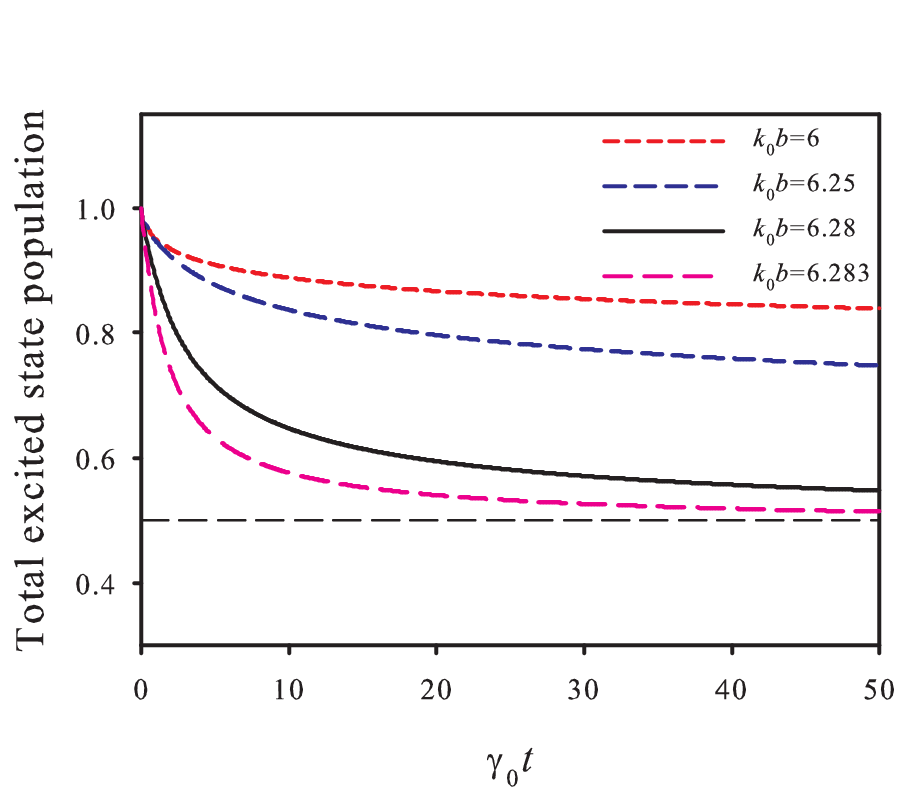}
\caption{\label{fig:four}
 Dynamics of the total excited state population. $k_0 a=3$. The coordinates of initially excited atom are $k_0 x_{exc}=0.8$, $k_0 y_{exc}=1.8$, $k_0 z_{exc}=0$. At $t=0$, only one Zeeman sublevel, $m_J = -1$, of the excited atom is populated. Random atomic ensemble occupy the spatial region $-L/2<z<L/2$ ($L$ is the length of the atomic ensemble). Atomic density is $n k_{0}^{-3}=0.002$, the number of atoms is $N=40$. The result is averaged over random spatial configurations of the ensemble using the Monte Carlo method. Dashed horizontal line refers the value 1/2.}\label{f4}
\end{figure}

Figure \ref{fig:four} refers to the atomic ensemble. It shows the dynamics of the total excited state population, i.e the sum of excited state probabilities of all the atoms. The curves shown in Fig. \ref{fig:four} were obtained by averaging the results over random spatial configurations of the ensemble using the Monte Carlo method. Here, one can clearly see that approaching the transverse size of a waveguide, $k_0 b$, to its critical value, $k_0 b=2\pi$, essentially changes the dynamics. It confirms the impact of evanescent modes on the character of many-body cooperative effects. Note that the curves in Fig. \ref{fig:four} do not decrease down to zero at the considered time interval. It is caused by the effect of incomplete spontaneous decay of atomic excitation in a single-mode waveguide \cite{13}.

\subsection{Steady-state transmission}
In this section we focus our attention on the atomic ensemble located in a single-mode waveguide. The ensemble is illuminated by external monochromatic probe radiation with the frequency $\omega_s$. The detuning of the probe frequency from exact resonance of a free atom is denoted by $\delta=\omega_s-\omega_0$. One of the sizes of the cross section is fixed, $k_0 a=3$, hereafter.

\begin{figure}\center
	\includegraphics[width=7cm]{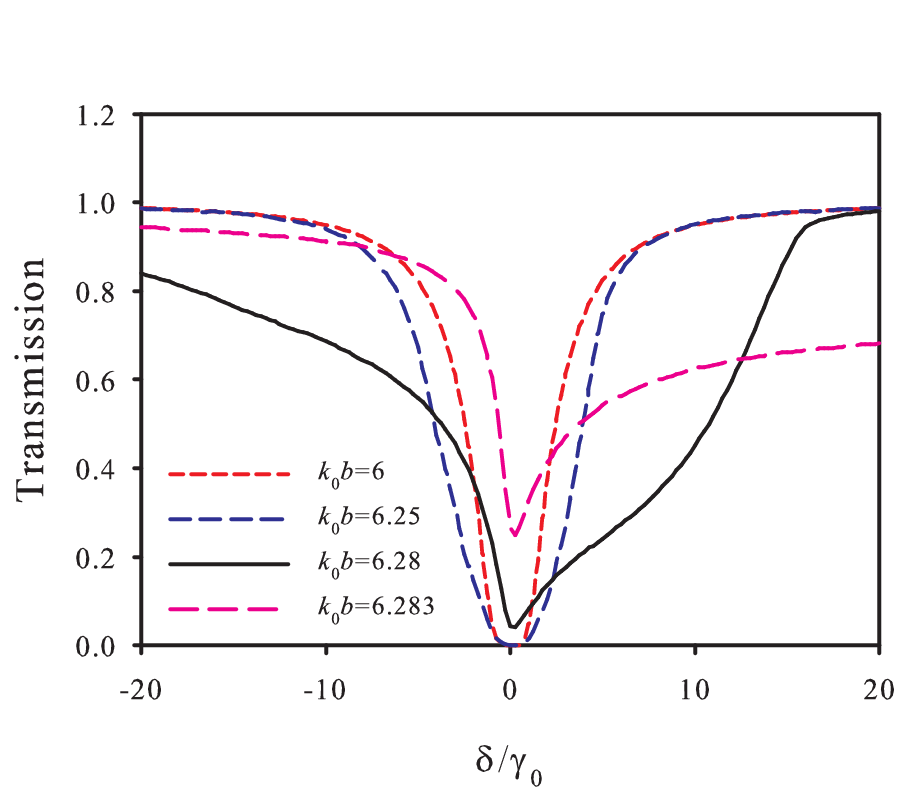}
\caption{\label{fig:five}
 The dependence of the transmission coefficient on the frequency of probe radiation. $\delta=\omega_s-\omega_0$ means the detuning of the probe frequency from the atomic resonance. $k_0 a=3$. Atomic density is $n k_{0}^{-3}=0.002$. $k_{0}L=1000$.}\label{f5}
\end{figure}

Figure \ref{fig:five} shows the dependence of the transmission coefficient on the frequency of probe radiation. In this figure, one can see the distortion of the spectrum shape for the values  of the parameter $k_0 b$ close to its critical value, $k_0 b=2\pi$. With approaching this critical value, the asymmetry of the profile appears, as well as the magnitude of the transmittance changes. Herewith, these changes manifest themselves differently for different frequencies of the probe radiation. In the case of resonant probe radiation, $\delta=0$, we observe monotonic increase of the transmission coefficient with the increase of the parameter $k_0 b$. The medium transforms from being completely reflective to being relatively transparent. For non-resonant radiation, the dependence of the transmission coefficient on the parameter $k_0 b$ may not be monotonic. In this case, evanescent modes can both weaken and enhance transmission.

\begin{figure}\center
	\includegraphics[width=7cm]{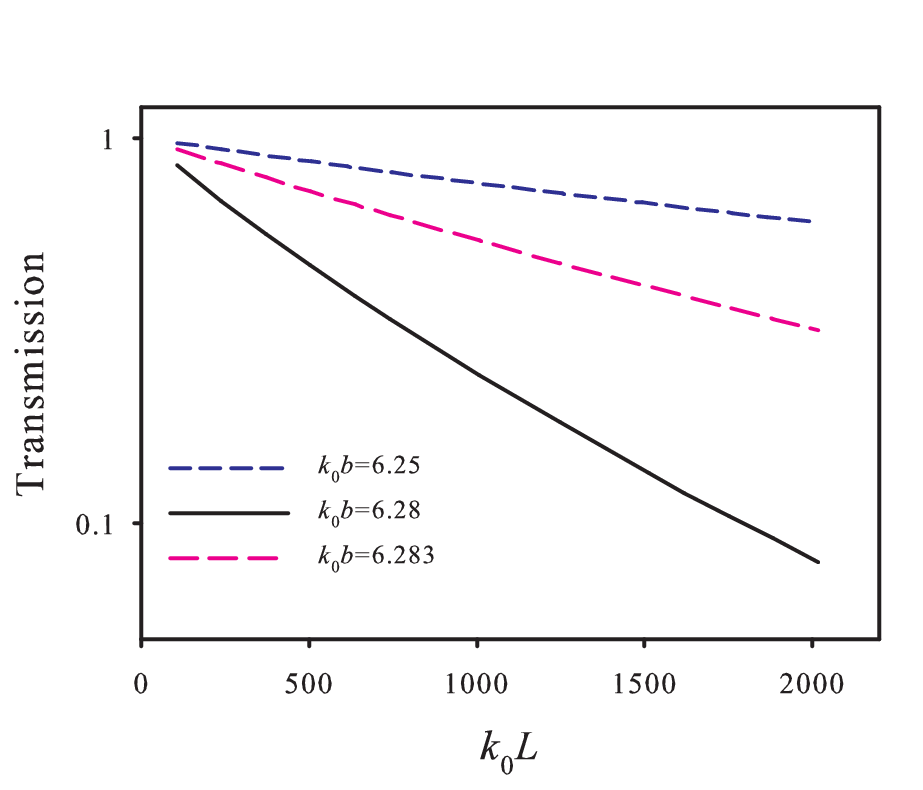}
\caption{\label{fig:six}
 Dependence of the transmission coefficient on the length of atomic ensemble. Probe frequency detuning is $\delta=5\gamma_0$. $k_0 a=3$. Atomic density is $n k_{0}^{-3}=0.002$.}\label{f6}
\end{figure}

The character of radiation transfer is exponentially decaying even for non-resonant probe radiation. This is illustrated by Fig. \ref{fig:six}. Such dependence of the transmission on the length of a sample is a signature of Anderson localization of light. Note that the length of localization nonmonotonically changes with the increase of the size $b$. When $k_0 b=6.25$, localization length equals to approx. $4\times 10^{3} k_{0}^{-1}$, that exceeds the length of atomic ensemble. When $k_0 b=6.28$, it equals to approx. $0.75\times 10^{3} k_{0}^{-1}$. When $k_0 b=6.283$, it equals to approx. $1.75\times 10^{3} k_{0}^{-1}$.

\begin{figure}\center
	\includegraphics[width=7cm]{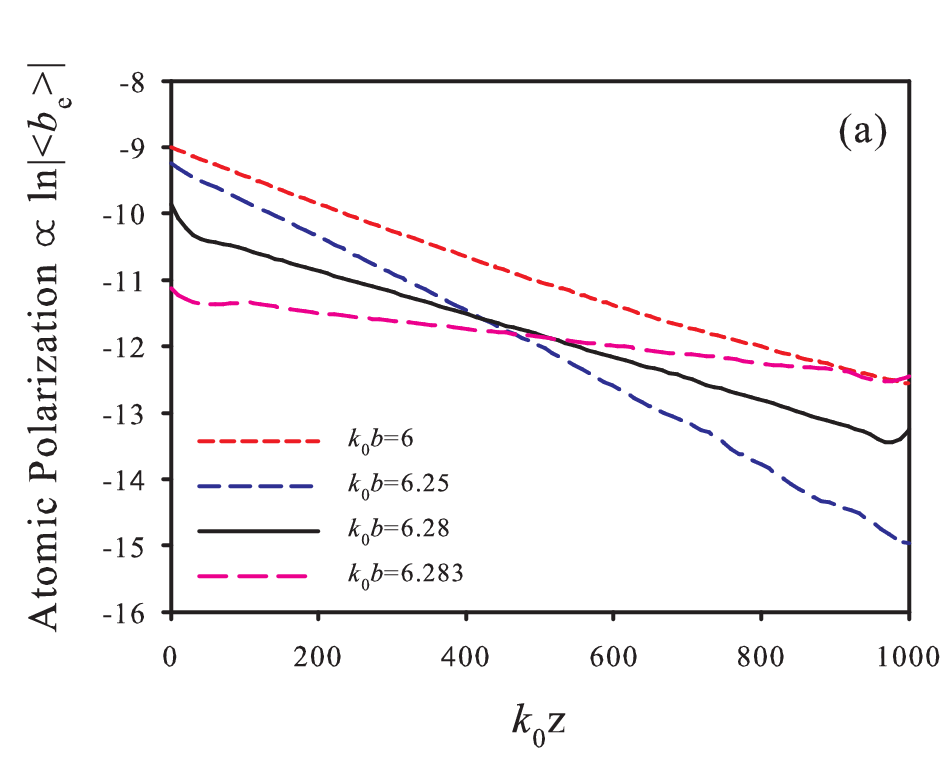}
	\includegraphics[width=7cm]{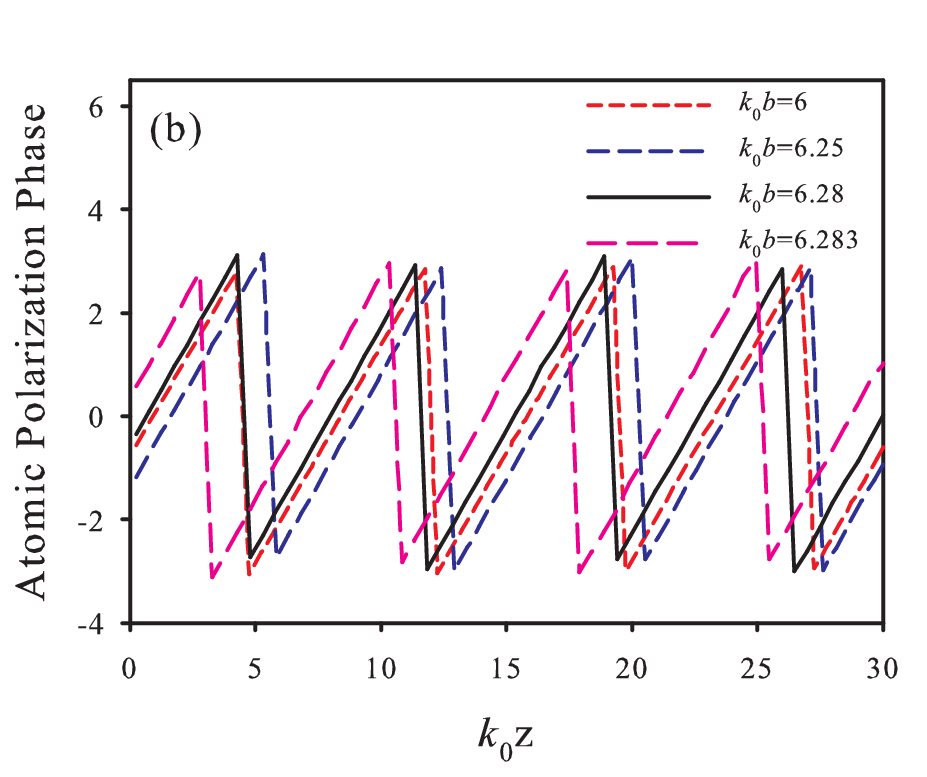}\\
	\caption{\label{fig:seven}
			Spatial distribution of the amplitude (a) and the phase (b) of the atomic polarization. $k_0 a=3$. Atomic density is $n k_{0}^{-3}=0.002$. Probe frequency detuning is $\delta=\gamma_0$. $k_{0}L=1000$.}\label{f7}
\end{figure}

Figure \ref{fig:seven} demonstrates the distribution of the coherent component of radiation inside the ensemble, which is proportional to the atomic polarization. Fig. \ref{fig:seven}(a) illustrates the absolute value of the atomic polarization depending of $z$ coordinate, Fig. \ref{fig:seven}(b) illustrates its phase. We observe that under considered conditions, the distribution of atomic polarization matches the Bouguer-Lambert-Beer law.

The slope of the lines shown in Fig. \ref{fig:seven}(a) determine the extinction coefficient. The performed analysis has shown that the extinction coefficient significantly depends on the sizes of the cross section of a waveguide when approaching them to critical values. With increasing the parameter $k_0 b$ from 6.28 to 6.283, i.e., by $0.05\%$, the extinction coefficient changes from 0.0032 to 0.0011, i.e., almost three times.

The slope of the lines shown in Fig. \ref{fig:seven}(b) determine the wavelength of light inside the medium and associated phase velocity. We observe that it almost does not depend on the parameter $k_0 b$. We explain such behavior by the fact that phase velocity is determined only by the radiation mode $\text{TE}_{01}$, and evanescent mode $\text{TE}_{02}$ almost does not affect it.

In the case of a single-mode waveguide, evanescent modes also manifest themselves in a significant population of Zeeman sublevel $m_{J}=0$ of the excited state. This sublevel cannot be populated by traveling wave $\text{TE}_{01}$. So its population is connected only with the dipole-dipole interaction, which is altered by exponentially decaying modes.

\begin{figure}\center
	\includegraphics[width=7cm]{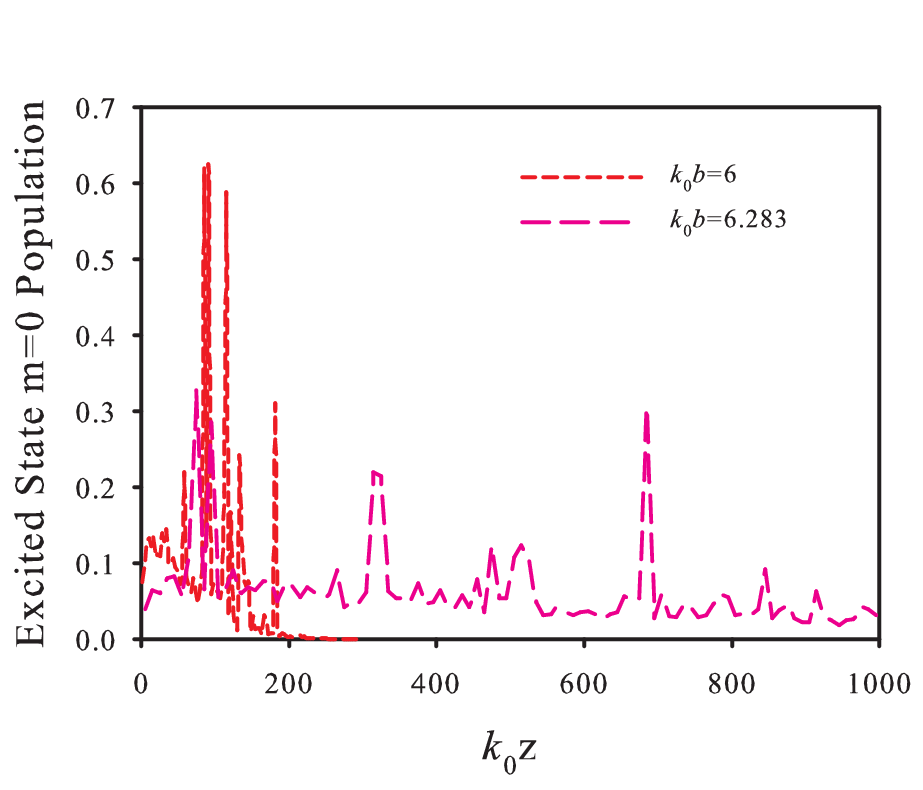}
\caption{\label{fig:eight}
 Spatial distribution of the population of the $|J=1, m_{J}=0\rangle$ state. $k_0 a=3$, $n k_{0}^{-3}=0.002$, $\delta=0$, $k_{0}L=1000$.}\label{f8}
\end{figure}

Figure \ref{fig:eight} shows the spatial distribution of the population of $|J=1, m_{J}=0\rangle$ state. The result is presented for both cases when the parameter $k_{0}b$ is far from its critical value, $k_{0}b=6$, and when it is close to the critical value, $k_{0}b=6.283$. In the first case, the sublevel $m_{J}=0$ of the excited state is populated only for atoms located near the front edge of the ensemble. This result is in the agreement with previously found fact that, in a single-mode waveguide, probe radiation is effectively reflected from the atoms closest to the source \cite{34}. Herewith, the spatial scale of the population distribution of the $|J=1, m_{J}=0\rangle$ state is determined by random position of the atom closest to the front edge and the next one. In the second case, when $k_{0}b=6.283$, the attenuation constant of the evanescent mode $\text{TE}_{02}$ is large, it exceeds the average interatomic distance more than 4 times. It results in the fact that we observe the distribution of $|J=1, m_{J}=0\rangle$ population close to uniform. Strong fluctuations in population density are noticeable, despite the fact that the number of Monte Carlo trials was $5\times 10^{6}$.

\begin{figure}\center
	\includegraphics[width=7cm]{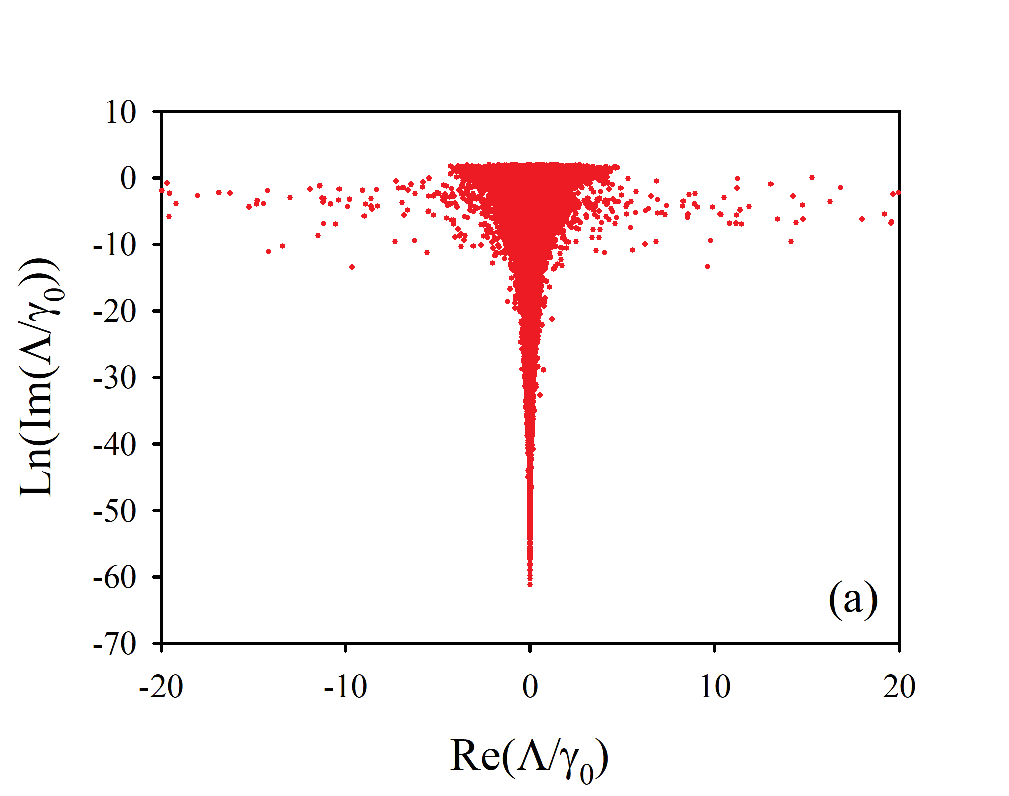}
	\includegraphics[width=7cm]{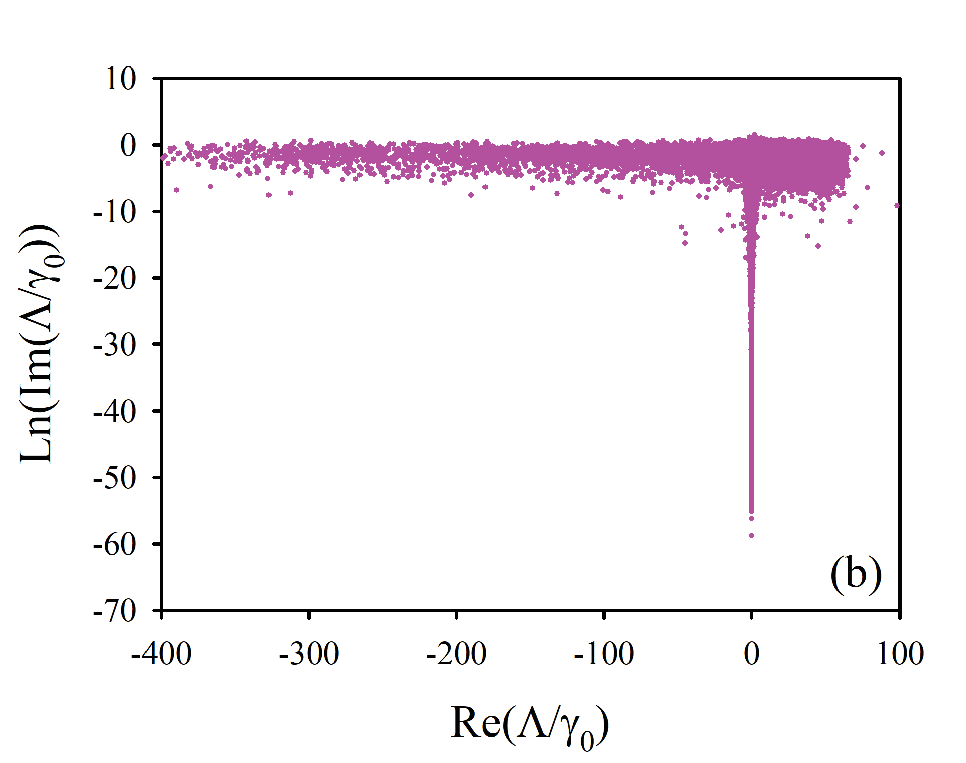}\\
\includegraphics[width=7cm]{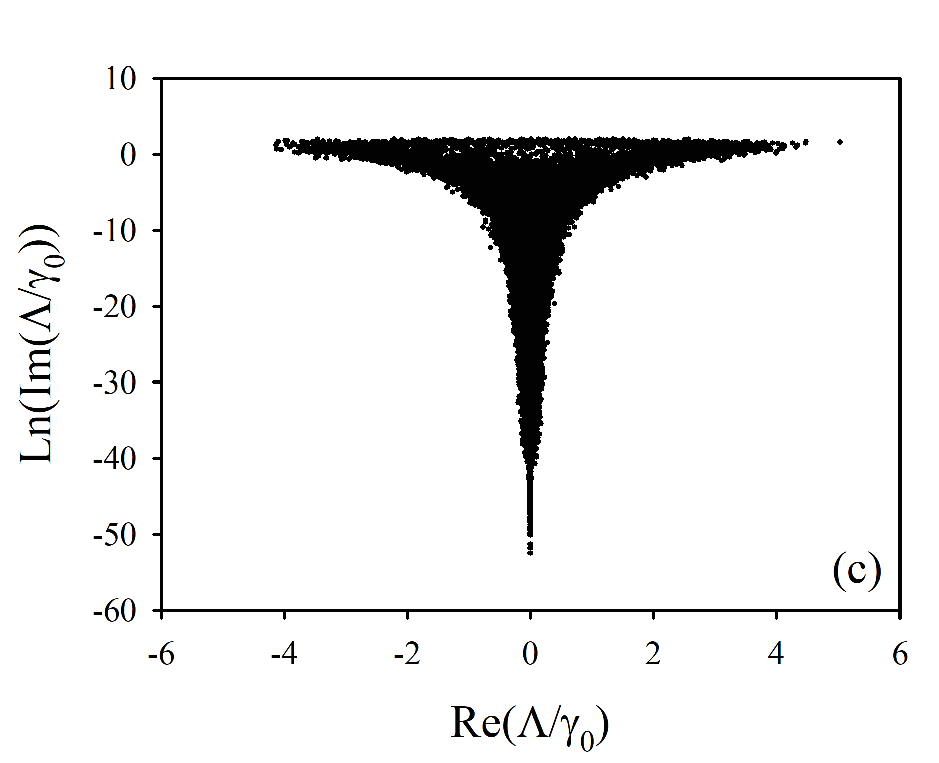}\\
	\caption{\label{fig:nine}
			The spectrum of the collective eigenstates of the atomic ensemble for two values of the transverse size of the waveguide: $k_{0}b=6$ -- (a) and $k_{0}b=6.283$ -- (b). For comparison, the spectrum of states without considering evanescent modes is shown -- (c). In the latter case, the spectrum does not change with changes in the transverse size. $k_0 a=3$, $n k_{0}^{-3}=0.002$, $k_{0}L=1000$.}\label{f9}
\end{figure}

The effects found in this work can be explained based on the analysis of changes in the spectrum of collective states with the change in the transverse size of a waveguide, approaching the parameter $k_{0}b$ to its critical value. The spectrum of collective states is a key parameter, which determines the optical properties of the atomic ensemble. Figure \ref{fig:nine} demonstrates the spectrum of collective states for two different values of the transverse size of a waveguide: $k_{0}b=6$ -- Fig. \ref{fig:nine}(a) and $k_{0}b=6.283$ -- Fig. \ref{fig:nine}(b). This scatter plot illustrates the distribution of the eigenvalues, $\Lambda$, of the matrix of re-emission of the atomic ensemble. Real part of $\Lambda$ determines the resonant frequency of the given collective state (cooperative Lamb shift). Imaginary part determines the linewidth of this collective state and, consequently, its lifetime. For comparison, we added the result obtained when evanescent modes are artificially eliminated -- Fig. \ref{fig:nine}(c). In the latter case, the picture is the same for both sizes $k_{0}b=6$ and $k_{0}b=6.283$.

The difference between Fig. \ref{fig:nine}(a) and Fig. \ref{fig:nine}(c) is that Fig. \ref{fig:nine}(a) contains a small number of significantly shifted states at the left and right edges of the figure. This is the effect of evanescent modes. In Fig. \ref{fig:nine}(b) one can clearly see tremendous broadening and asymmetry of the spectrum of collective states. This explains the distortion of the profiles observed in Fig. \ref{fig:five} for the case when the parameter $k_{0}b$ approaches its critical value, $k_{0}b=2\pi$.

\section{Conclusion}
We have performed a comprehensive study of the influence of evanescent modes of the electromagnetic field in a waveguide on the character of interatomic dipole-dipole interaction and associated cooperative effects. This influence has been investigated based on the analysis, how radiative and spectral properties of an atomic ensemble are modified with the change of the transverse sizes of the waveguide approaching them to their critical values.

As a proof of concept, we have demonstrated that evanescent mode can couple far separated atoms in a zero-mode waveguide. In the case of a single-mode waveguide, evanescent mode significantly modifies the long-range dipole-dipole interaction when the transverse size of a waveguide approaches its critical value.

The main focus of this work is given to the analysis of the transmittance of an atomic ensemble located in a single-mode waveguide in the steady-state regime. When analyzing the dependence of the transmission coefficient on the probe frequency, we have revealed the asymmetry of the profile. This asymmetry is significantly enhanced when the size of a waveguide approaches its critical value, $k_{0}b=2\pi$. It is accompanied by the change of the absolute value of the transmission.
The character of radiation transfer is found to be exponentially decaying even for non-resonant probe, i.e. Anderson localization of light takes place. The length of localization significantly changes with the change of the transverse size in the vicinity of the critical value, and this dependence has complex nonmonotonic character.

Additionally, we have analyzed the propagation of the coherent component of radiation inside the atomic ensemble in a single-mode waveguide. On this basis, the extinction coefficient and phase velocity were calculated. We have shown that the extinction coefficient is very sensitive to the size of the cross section of a waveguide when it approaches to its critical value. It is connected with the effect of the evanescent mode. However, phase velocity is determined only by the radiation mode, so small changes of the transverse size of a waveguide near its critical value almost does not affect it.

For better understanding of the peculiarities of cooperative effects caused by strong manifestation of evanescent modes, we have analyzed the spectrum of eigenstates of atomic ensemble located in a single-mode waveguide whose transverse size is close to the critical value. It was established that the spectrum of collective eigenstates is tremendously broadened and asymmetric.

\section*{Acknowledgments}
A.S.K. appreciates financial support from the
Foundation for the Advancement of Theoretical Physics and Mathematics "BASIS" (Agreement 24-1-3-12-1). I.M.S. acknowledges the financial support of the Ioffe Institute  within the framework of the baseline project FFUG-2024-0039. The results of the work were obtained using computational resources of Peter the Great St. Petersburg Polytechnic University Supercomputer
Center (https://scc.spbstu.ru/).


\vspace*{-5mm}

\begin{references}
\bibitem{1} M. Gross and S. Haroche, Phys. Rep. \textbf{93}, 301 (1982).
\bibitem{2} A. Gonzalez-Tudela and D. Porras, Phys. Rev. Lett. \textbf{110}, 080502 (2013).
\bibitem{3} S. Mokhlespour, J. E. M. Haverkort, G. Slepyan, S. Maksimenko, and A. Hoffmann, Phys. Rev. B \textbf{86}, 245322 (2012).
\bibitem{4}  A. I. Galimov, M. V. Rakhlin, G. V. Klimko, Yu. M. Zadiranov, Yu. A. Guseva, S. I. Troshkov, T. V. Shubina, and A. A. Toropov,  Jetp Lett. \textbf{113}, 252 (2021).
\bibitem{5} S. A. Moiseev, M. M. Minnegaliev, K. I. Gerasimov, E. S. Moiseev, A. D. Deev, Yu. Yu. Balega, Phys. Usp. \textbf{68},  431 (2025).
\bibitem{6} D. D. Awschalom, R. Hanson, J. Wrachtrup,  and B. B. Zhou, Nature Photonics \textbf{12}, 516 (2018).
\bibitem{7} M. W. Doherty, N. B. Manson, P. Delaney, F. Jelezko, J. Wrachtrup, and L. C. Hollenberg, Phys. Rep. \textbf{528}, 1 (2013).
\bibitem{8} E. M. Purcell, Proceedings of the American Physical Society \textbf{69}, 681 (1946).
\bibitem{9} D. Kleppner, Phys. Rev. Lett. \textbf{47}, 233 (1981).
\bibitem{10} J. P. Dowling, M. O. Scully, and F. DeMartini, Opt. Comm. \textbf{82}, 415 (1991).
\bibitem{11} J. P. Dowling, Foundations of Physics \textbf{23}, 895 (1993).
\bibitem{12} G. S. Agarwal and S. D. Gupta, Phys. Rev. A \textbf{57}, 667 (1998).
\bibitem{13} A. S. Kuraptsev and I. M. Sokolov, Phys. Rev. A \textbf{101}, 053852 (2020).
\bibitem{14} G. S. Agarwal, Phys. Rev. A \textbf{12}, 1475 (1975).
\bibitem{15} D. Kleppner, Phys. Rev. Lett. \textbf{47}, 233 (1981).
\bibitem{16} V. V. Klimov and M. Ducloy, Phys. Rev. A \textbf{62}, 043818 (2000).
\bibitem{17} V. V. Klimov and M. Ducloy, Phys. Rev. A \textbf{69}, 013812 (2004).
\bibitem{18} V. I. Yudson and P. Reineker, Phys. Rev. A \textbf{78}, 052713 (2008).
\bibitem{19} V. V. Klimov, Phys. Usp. \textbf{64}, 990 (2021).
\bibitem{20} T. Kobayashi, Q. Zheng, and T. Sekiguchi, Phys. Rev. A \textbf{52}, 2835 (1995).
\bibitem{21} E. V. Goldstein and P. Meystre, Phys. Rev. A \textbf{56}, 5135 (1997).
 \bibitem{22} R. Rohlsberger, K. Schlage, B. Sahoo, S. Couet, and R. Ruffer, Science \textbf{328}, 1248 (2010).
\bibitem{23} Y.-Q. Zhang, L. Tan, and P. Barker, Phys. Rev. A \textbf{89}, 043838 (2014).
\bibitem{24} A. Goban, C.-L. Hung, J. D. Hood, S.-P. Yu, J. A. Muniz, O. Painter, H. J. Kimble, Phys. Rev. Lett. \textbf{115}, 063601 (2015).
\bibitem{25} A. S. Kuraptsev and I. M. Sokolov, J. Exp. Theor. Phys. \textbf{123}, 237 (2016).
\bibitem{26} A. S. Kuraptsev and I. M. Sokolov, Phys. Rev. A \textbf{94}, 022511 (2016).
\bibitem{27} M. D. Lee, S. D. Jenkins, Y. Bronstein, and J. Ruostekoski, Phys. Rev. A \textbf{96}, 023855 (2017).
\bibitem{28} T. Botzung, D. Hagenm\"{u}ller, S. Sch\"{u}tz, J. Dubail, G. Pupillo, and J. Schachenmayer, Phys. Rev. B \textbf{102}, 144202 (2020).
\bibitem {Bufetov1} A. D. Pryamikov, A. V. Gladyshev, A. F. Kosolapov, and I. A. Bufetov, Phys. Usp. \textbf{67}, 129 (2024).
\bibitem{Chabanov} A. A. Chabanov, M. Stoytchev, and A. Z. Genack, Nature \textbf{404}, 850 (2000).
\bibitem{Abmann}   M. Abmann and M. Bayer, Adv. Quantum Tech \textbf{3}, 1900134 (2020).
\bibitem{30} P. W. Anderson, Phys. Rev. \textbf{109}, 1492 (1958).
\bibitem{31} D. S. Wiersma, P. Bartolini, A. Lagendijk, and R. Righini, Nature \textbf{390}, 671 (1997).
\bibitem{32} M. Storzer, P. Gross, C. M. Aegerter, and G. Maret, Phys. Rev. Lett. \textbf{96}, 063904 (2006).
\bibitem{33} T. Sperling, W. Buhrer, C. M. Aegerter, and G. Maret, Nat. Photonics \textbf{7}, 48 (2013).
\bibitem{34} A. S. Kuraptsev and I. M. Sokolov,  Rev. A \textbf{105}, 063513 (2022).
\bibitem{nanofiber1} N. V. Corzo, B. Gouraud, A. Chandra A. Goban, A. S. Sheremet,
D. V. Kupriyanov, and J. Laurat, Phys. Rev. Lett. \textbf{117}, 133603 (2016).
\bibitem{nanofiber2} H. L. Sorensen, J.-B. Beguin, K. W. Kluge, I. Iakoupov, A. S. Sorensen, J. H. Muller, E. S. Polzik, and J. Appel, Phys. Rev. Lett. \textbf{117},
133604 (2016).
\bibitem {Naumov2} A. V. Naumov, A. A. Gorshelev, M. G. Gladush, T. A. Anikushina, A. V. Golovanova, J. Kohler, and L. Kador, Nano Letters \textbf{18}, 6129 (2018).
\bibitem{Skipetrov2025} S. E Skipetrov and I. M. Sokolov, Phys. Rev. B \textbf{112}, 064206 (2025).
\bibitem{Rebane}   R. A. Avarmaa and K. K. Rebane,  Sov. Phys. Usp. \textbf{31}, 225 (1988).


\bibitem{F45} L. L. Foldy, Phys. Rev. \textbf{67}, 107 (1945).
\bibitem{L51} M. Lax,  Rev. Mod. Phys. \textbf{23}, 287 (1951).

\bibitem{Javanainen:1999}
J.~Javanainen, J.~Ruostekoski, B.~Vestergaard, and M.~R. Francis, Phys. Rev. A \textbf{59}, 649  (1999).

\bibitem{RMO00} M. Rusek, J. Mostowski, and A. Orlowski, Phys. Rev. A \textbf{61}, 022704 (2000).

\bibitem{34a} H. Fu and P. R. Berman, Phys. Rev. A \textbf{72}, 022104 (2005).

\bibitem{Svidzinsky:2010}
A.~A. Svidzinsky, J. T.~Chang, and M.~O. Scully,   Phys. Rev. A \textbf{81}, 053821 (2010).
\bibitem{KSH11}
I.~M. Sokolov, D.~V. Kupriyanov, and M.~D. Havey,  J. Exp. Theor. Phys. \textbf{112}, 246 (2011).

\bibitem{38a} D. V. Kuznetsov, Vl. K. Rerikh and M. G. Gladush, J. Exp. Theor. Phys. \textbf{113}, 647 (2011).

\bibitem{SS14} S. E. Skipetrov and I. M. Sokolov, Phys. Rev. Lett. \textbf{112}, 023905 (2014).

\bibitem{BGAK14} L. Bellando, A. Gero, E. Akkermans, and R. Kaiser,  Phys. Rev. A \textbf{90}, 063822 (2014).

\bibitem{KS_PRA2015} A. S. Kuraptsev and I. M. Sokolov, Phys. Rev. A \textbf{91}, 053822
(2015).

\bibitem{Guerin:2016a}
W.~Guerin, M.~O. Ara\'ujo, and R.~Kaiser,  Phys. Rev. Lett. \textbf{116}, 083601 (2016).



\bibitem{31a} A. B. Shesterikov, M. Y. Gubin, M. G. Gladush, and A. V. Prokhorov, J. Exp. Theor. Phys. \textbf{124}, 18 (2017).

\bibitem{Guerin:2017a}
W.~Guerin, M.~T. Rouabah, and R.~Kaiser,  J. Mod. Opt.   \textbf{64}, 895 (2017).

\bibitem{Kuraptsev:2017} A.~S. Kuraptsev, I.~M. Sokolov, and M.~D. Havey,  Phys. Rev.   A \textbf{96}, 023830 (2017).

\bibitem{Skipetrov_2019b} S. E. Skipetrov and I. M. Sokolov, Phys. Rev. Lett. \textbf{123}, 233903  (2019).

\bibitem{35}A. S. Kuraptsev and I. M. Sokolov, Phys. Rev. A \textbf{107}, 042808 (2023).
\bibitem{36}J. D. Jackson, Classical Electrodynamics (Wiley, New York, 1962).
\bibitem{37} W. R. Raudorf, American Journal of Physics \textbf{46}, 35 (1978).




\end{references}
\end{document}